\documentclass[12pt,preprint,natbib]{aastex6}

\begin{document}

\title{Clarifying the Status of HD 100546 as Observed by the Gemini Planet Imager}
\author{Thayne Currie}
\affiliation{National Astronomical Observatory of Japan, Hilo, Hawaii, USA}
\affiliation{Eureka Scientific
2452 Delmer Street Suite 100
Oakland, California 94602 USA}
\author{Sean Brittain}
\affiliation{Department of Physics \& Astronomy, 118 Kinard Laboratory, Clemson University, Clemson, SC 29634 USA}
\author{Carol A. Grady}
\affiliation{Stellar Astrophysics and Exoplanets Laboratory, Code 667, NASA-Goddard Space Flight Center, Greenbelt, MD USA}
\affiliation{Eureka Scientific
2452 Delmer Street Suite 100
Oakland, California 94602 USA}
\author{Scott J. Kenyon}
\affiliation{Smithsonian Astrophysical Observatory, Cambridge, Massachusetts USA}
\author{Takayuki Muto}
\affiliation{Division of Liberal Arts, Kogakuin University, Tokyo, Japan}
\newpage
\newpage
\keywords{planetary systems, stars: early-type, stars: individual: HD 100546}
\section{Introduction}
 \textbf{HD 100546} is a young, early-type star and key laboratory for studying gas giant planet formation.
 Using the \textit{Gemini Planet Imager},
  \citet{Currie et al. 2015} reported the near-infrared (IR) detection of the thermal IR-discovered protoplanet candidate HD 100546 b \citep{Quanz et al. 2013}, arguing that its colors may reveal circumplanetary material.   They also identified an emission source at 13 au:  a weakly-polarized disk feature or a possible 2nd candidate \citep[HD 100546 "c",][]{Brittain et al. 2014} responsible for HD 100546's inner disk cavity and thermal IR-bright spiral \citep{Currie et al. 2014,Quanz et al. 2015}.  
Recently, two papers from the GPI Campaign Team (GPIES) presenting newer 2016 data argue that the detections of HD 100546 b and "c" in the 2015 GPI data are artifacts of spatial filtering and/or aggressive processing.
 
 \section{Results}
\begin{figure*}[h]
\includegraphics[angle=270,scale=0.65]{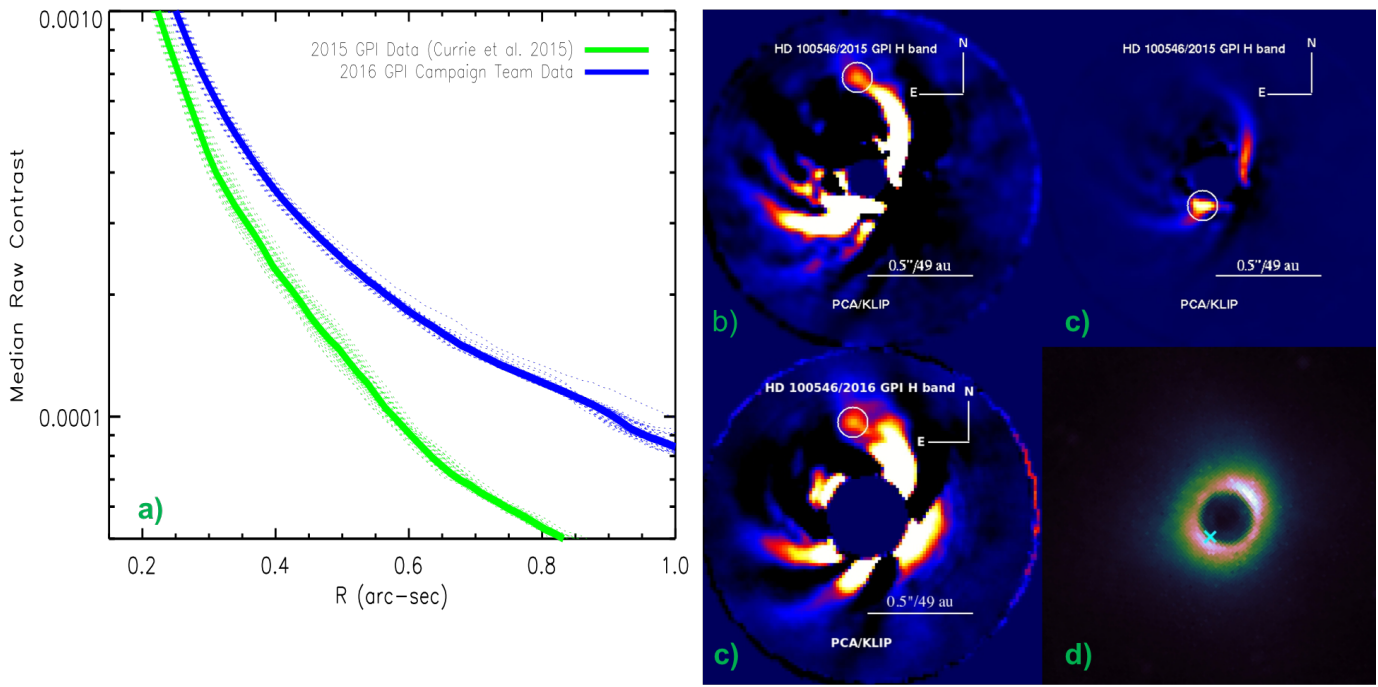}
\caption{a) The  2015 GPI data have fainter halo intensity profiles than the 2016 GPIES data and constitute higher-quality data;  b) HD 100546 b and c) the emission source HD 100546 ``c" detected in the 2015 data using KLIP, removing one principal component per annulus and imposing a 0.5 $\lambda$/D rotation gap (S/N $\sim$ 7 for both).   d) Detection of HD 100546 b from 2016 GPIES $H$-band data using KLIP, removing 5 principal components per annulus and using 1.5 $\lambda$/D rotation gap (S/N $\sim$ 4.5--5).    
Its position is slightly offset between 2015 and 2016.
  e) Position of the HD 100546 ``c" emission source (cross) in a representative 2016 GPIES data cube from \citet{Follette et al. 2017}, lying largely underneath the coronagraph mask (dark region), unlike in the 2015 data.  \label{fig1} }
\end{figure*}

 Our independent, detailed analyses of the 2016 GPIES data and rereduction of 2015 data using different methods support the findings of \citet{Currie et al. 2015}.  Given the strong, active interest in HD 100546, we summarize some of these findings below and in Figure \ref{fig1}:
 
   \begin{itemize}
 \item The  2015 \citeauthor{Currie et al. 2015} data are of much higher quality than the 2016 GPIES data and should be given priority in parsing HD 100546's complex environment. 
 The stellar halo from the 2015 data is less variable and ranges between 25\% and 300\% fainter than in the 2016 GPIES data.   
  The typical wavefront errors recorded in the GPI fits headers for the 2015 data are 30\% lower ($\approx$ 110 nm vs. 145 nm).
 
 \item Using KLIP \citep[][]{Soummer et al. 2012}, we reaffirm the detection of HD 100546 b and the putative ``c" object in a rereduction of our 2015 data.    
  As we do not use spatial filtering now or in our previous paper, HD 100546 b's point source-like appearance is not a filtering artifact as suggested in \citet{Rameau et al. 2017}.   We also detect HD 100546 b in the 2016 GPIES data.   The GPIES reduction approaches are not well suited for detecting either object as they (with their KLIP approach) sub-optimally model the stellar point-spread function/lack a rotation gap or \citep[with LOCI;][]{Lafreniere et al. 2007} allow the disk's bright, spatially-varying emission to contaminate the algorithm's covariance matrix.  
  Our A-LOCI \citep{Currie et al. 2012} and KLIP reductions have high throughput ($>$ 70\%--80\% for 'b').   The detections of ``c" are not due to aggressive processing as proposed in \citet{Follette et al. 2017}.   Instead, aggressive settings (i.e. a full-rank LOCI) \textit{preclude} robust detections of either source and much of the disk.  Rigorous forward-modeling employed in \citet{Currie et al. 2015} but not considered in \citet{Follette et al. 2017} provided evidence that ``c" is a real astrophysical feature even if the interpretation (disk feature or protoplanet) is unclear.

 \item \citet{Rameau et al. 2017} proposed that HD 100546 b's spectrum resembles the star's and the outer disk's and thus simply reveals scattered-light disk emission.   However, HD 100546 has sub-au scale, optically-thin near-IR excess emission ($J$-$K$ $\sim$ 1, over 50\% of the system's $H$ band flux) unresolvable by GPI \citep{Benisty2010, Tatulli2011}: the GPI stellar spectrum shown in \citet{Rameau et al. 2017} is instead that of the star+inner disk.    
Furthermore, the light scattered by the outer disk is dominated by the 1500--1750 K inner disk; the disk comparison regions used by GPIES also lie on a thermal IR-bright spiral, which (if driven by a massive planet) can be heated well above local temperature \citep[$\sim$ 450 K;][]{Lyra et al. 2016}.  HD 100546 b's temperature is likely intermediate between these two values \citep{Quanz et al. 2015}: distinguishing between scattered inner disk light and thermal emission from a protoplanet likely requires multi-wavelength data.
  
\item The position of the coronagraph in the 2016 GPIES data heavily attenuates any source at HD 100546 "c"'s position.   A 2016 detection may also be precluded if "c" has passed behind the disk wall \citep[see][]{Currie et al. 2015}.

 \end{itemize}

New, high-quality GPI data can further explore HD 100546's circumstellar environment.  While poor weather at Cerro Pachon has impeded our follow-up observations over the past 2.5 years, we will present a comprehensive analysis of HD 100546 with GPI once our program is completed.

\acknowledgments
We thank Bruce Macintosh for helpful comments.


\begin{thebibliography}{}
\bibitem[Benisty et al.(2010)]{Benisty2010}Benisty, M., Tatulli, E., M\'enard, F., Swain, M. R., 2010, A\&A, 511, 75
\bibitem[Brittain et al.(2014)]{Brittain et al. 2014}Brittain, S., Carr, J. S., Najita, J. R., et al., 2014, \apj, 791, 136
\bibitem[Currie et al.(2012)]{Currie et al. 2012}Currie, T., Debes, J. H., Rodigas, T. J., et al., 2012, \apj, 760, L32
\bibitem[Currie et al.(2014)]{Currie et al. 2014} Currie, T., Muto, T., Kudo, T., et al., 2014, \apj, 796, L30
\bibitem[Currie et al.(2015)]{Currie et al. 2015}Currie, T., Cloutier, R., Brittain, S., et al., 2015, \apj, 814, L27
\bibitem[Follette et al.(2017)]{Follette et al. 2017}Follette, K., B., Rameau, J., Dong, R., et al., 2017, \aj, 153, 264
\bibitem[Lafreni\`ere et al.(2007)]{Lafreniere et al. 2007}Lafreni\'ere, D., Marois, C., Duyon, R., et al., 2007, \apj, 660, 770
\bibitem[Lyra et al.(2016)]{Lyra et al. 2016}Lyra, W., Richert, A., Boley, A., et al., 2016, \apj, 817, 102
\bibitem[Quanz et al.(2013)]{Quanz et al. 2013}Quanz, S., Meyer, M. R., Kenworthy, M., et al., 2013, \apj, 766, L1
\bibitem[Quanz et al.(2015)]{Quanz et al. 2015}Quanz, S., Amara, A., Meyer, M. R., et al., 2015, \apj, 807, 64
\bibitem[Rameau et al.(2017)]{Rameau et al. 2017}Rameau, J., Follette, K. B., Pueyo, L., et al., 2017, \aj, 153, 244
\bibitem[Soummer et al.(2012)]{Soummer et al. 2012}Soummer, R., Pueyo, L., Larkin, J.,  2012, \apj, 755, L28
\bibitem[Tatulli et al.(2011)]{Tatulli2011}Tatulli, E., Benisty, M., Menard, F., et al., 2011, A\&A, 531, 1
\end{thebibliography}
\end{document}